\documentclass[12pt]{article}
\usepackage{epsf}
\usepackage{color}
\usepackage{graphicx}
\usepackage{amsmath}
\usepackage{amssymb}
\usepackage{latexsym}

\newcommand{\Gc}[2]{\dimen0=\ht\strutbox
                    \advance\dimen0\dp\strutbox
                    \multiply\dimen0 by#1
                    \divide\dimen0 by2
                    \advance\dimen0 by-.5\normalbaselineskip
                    \raisebox{-\dimen0}[0pt][0pt]{#2}}

\begin{document}

\newcommand{\mathcalpl}{M_{\mathcalathrm{Pl}}}
\setlength{\baselineskip}{18pt}
\begin{titlepage}
\begin{flushright}
OU-HET 667/2010 
\end{flushright}

\vspace*{1.2cm}
\begin{center}
{\Large\bf Phenomenological Aspects of \\ 
Dirichlet Higgs Model from Extra-Dimension}
\end{center}
\lineskip .75em
\vskip 1.5cm

\begin{center}
{\large 
Naoyuki Haba$^{a,}$\footnote[1]{E-mail:
\tt haba@phys.sci.osaka-u.ac.jp}, 
Kin-ya Oda$^{a,}$\footnote[2]{E-mail:
\tt odakin@phys.sci.osaka-u.ac.jp}, 
and 
Ryo Takahashi$^{b,}$\footnote[3]{E-mail: 
\tt ryo.takahashi@mpi-hd.mpg.de}
}\\

\vspace{1cm}

$^a${\it Department of Physics, Graduate School of Science, Osaka
University, \\
Toyonaka, Osaka 560-0043, Japan}\\
$^b${\it Max-Planck-Institut f$\ddot{u}$r Kernphysik, Postfach 10 
39 80, 69029 \\ Heidelberg, Germany}\\

\vspace*{10mm}
{\bf Abstract}\\[5mm]
{\parbox{13cm}{
We study a simple five-dimensional extension of the Standard Model,
 compactified on a flat line segment in which there propagate
 Higgs and gauge bosons of the Standard Model. 
We impose a Dirichlet boundary condition on the Higgs field to realize 
 its vacuum expectation value.
Since a flat Nambu-Goldstone zero-mode of the bulk Higgs is
 eliminated by the Dirichlet boundary condition, 
 a superposition of the Higgs Kaluza-Klein modes 
 play the role of the Nambu-Goldstone boson except at the boundaries.
We discuss phenomenology of our model at the
 LHC, namely the top Yukawa deviation and the production and
 decay of the physical Higgs field, as well as the 
constraints from the electroweak precision measurements.
}}
\end{center}
\end{titlepage}

\section{Introduction}

Extra-dimensional theory is an interesting candidate beyond the Standard Model (SM), having rich 
phenomenology of Kaluza-Klein (KK) particles, a tower of modes for each SM field propagating in 
bulk. Especially, in the Universal Extra Dimensions (UED) model~\cite{ued}, the lightest KK 
particles (LKP) with an odd parity is stable and can be a candidate for the dark matter. The UED 
can also lower constraints on the compactification scale from the electroweak (EW) precision 
measurements to few hundred GeV due to the five-dimensional (5D)  Lorentz symmetry 
\cite{Appelquist,Gogoladze:2006br}, while the scale is restricted to be larger than few TeV for the
 brane localized fermion (BLF) scenario \cite{kk1}-\cite{Cheung:2001mq}. If the KK-modes are 
confirmed at the CERN LHC experiment, it would strongly indicate the existence of the extra 
dimensions.

On the other hand, anther interesting phenomenological consequence of the extra-dimensional theory 
is the top Yukawa deviation, which is a deviation of the Yukawa coupling between top quark and 
physical Higgs field from the naive SM expectation: $m_t/v_{\text{EW}}$. Such a deviation is generally
 caused when there are several Higgs doublets. Recently it has been pointed out that the deviation 
can be induced from effects of the brane localized Higgs potentials~\cite{Haba:2009uu,Haba:2009wa} 
and the Dirichlet boundary condition (BC)~\cite{Haba:2009pb} in a simple setup of the 
extra-dimensional theory with one Higgs doublet.\footnote{Yukawa deviation is considered in a 
warped gauge-Higgs unification model~\cite{Hosotani:2008by}, and a deviation between SM-field and 
flavon  is considered in~\cite{Holthausen:2009qj}.}

In this paper, we study a 5D model~\cite{Haba:2009pb} where the gauge bosons and Higgs doublet of 
the SM exist in the 5D bulk compactified on a line segment with a flat metric, imposing a generic 
Dirichlet BC for the Higgs and Neumann BC for all other SM fields. The vacuum expectation value 
(VEV) of the Higgs is induced from the Dirichlet BC which is generally allowed in higher 
dimensional theories. We present wave-function profiles of bulk fields under these BCs. The 
Dirichlet BC removes the Higgs zero-mode and one might worry how a Nambu-Goldstone (NG) boson is 
supplied for the electroweak symmetry breaking. We show that a superposition of sufficiently large 
number of KK-modes gives a correct wave-function profile which is flat except at the boundaries.

We consider the case where all other SM fermions propagate in the bulk as the 
UED model. 
Then, we will give detailed phenomenology of 
the model on the top Yukawa deviation, the production and decay of Higgs, 
constraints on the model from the electroweak precision measurements, and dark 
matter. 

The organization of the paper is as follows. In the next section, we present our extra-dimensional 
setup and show the profiles, masses and couplings of bulk Higgs and gauge fields. In Section 3, we 
discuss phenomenological aspects of the setup such as top Yukawa deviation, the production and 
decay of the Higgs, constraints from the EW precision measurements, and dark matter candidate. In 
Section 4, we summarize our results. In Appendix we show the 5D gauge and Higgs kinetic Lagrangian 
and the gauge fixing.

\section{Bulk Higgs and its profiles}

Our setup is that the SM Higgs doublet exists in the 5D flat space-time with  Dirichlet BC. We 
study the Higgs wave-function profiles, masses and couplings at first. Our analysis is applicable 
in more general cases though we will focus on the case where EW symmetry is broken through the 
extra-dimensional BC.

\subsection{VEV profile}

Let us show wave-function profile of classical mode, i.e.\ the VEV, of a bulk scalar field. Start 
from an illustrative example of the bulk complex scalar $\Phi$ with the kinetic action
 \begin{eqnarray}
  S=-\int d^4x\int_{-L/2}^{+L/2}dz\,\left|\partial_M\Phi\right|^2,
 \end{eqnarray}
where we write 5D coordinates as $x^M=(x^\mu,z)$ with $M,N,\ldots$ running from 0 to 4 and 
$\mu,\nu,\ldots$ running from 0 to 3. The 5D spacetime is compactified on a line segment and the 
action is defined in $-L/2\leq z \leq+L/2$. 

Now let us impose a Dirichlet BC for the bulk scalar field. We consider the action, along with all 
the BCs, is invariant under the KK parity, $z\to-z$. A compactification on the line segment with 
the Dirichlet BC~\cite{Haba:2009pb} derives different physics from the orbifold $S^1/Z_2$ 
compactification with a brane-localized potential in an infinitely large coupling limit 
\cite{Haba:2009uu}, since the former does not have the uncontrollably large interactions involving 
the NG-boson as we will see.

Rewriting in terms of real fields, $\Phi=(\Phi_R+i\Phi_I)/\sqrt{2}$, the variation of the action 
reads
 \begin{eqnarray}
  \delta S=\int d^4x\int_{-L/2}^{+L/2}dz\bigg[\delta\Phi_X(\mathcal{P}\Phi_X)
           +\delta(z-\frac{L}{2})\delta\Phi_X(-\partial_z\Phi_X)
           +\delta(z+\frac{L}{2})\delta\Phi_X(+\partial_z\Phi_X)\bigg],
 \end{eqnarray}
where $\mathcal{P}\equiv\Box+\partial_z^2$. The VEV of the scalar field $\Phi^c$ is determined by 
the action principle $\delta S=0$, that is, $\mathcal{P}\Phi_X^c=0$. The general solution of this 
Equation of Motion (EoM) is $\Phi^c(z)=A+Bz$. The undetermined coefficients $A$ and $B$ can be 
fixed by choosing BCs at $z=\pm L/2$. We have four choices of combination of Dirichlet and Neumann 
BCs at $z=\pm L/2$, namely, 
 \begin{eqnarray}
  (D,D),~~~(D,N),~~~(N,D),~~~\mbox{and}~~~(N,N),
 \end{eqnarray}
where $D$ and $N$ denote the Dirichlet and Neumann BCs, respectively. These BCs are written as
 \begin{eqnarray}
  \delta\Phi|_{z=\xi}=0,
 \end{eqnarray}
for the Dirichlet BC, and 
 \begin{eqnarray}
  \partial_z\Phi|_{z=\xi}=0,
 \end{eqnarray}
for the Neumann BC, where $\xi$ is taken as $+L/2$ or $-L/2$ in each case. Difference choice of BC 
corresponds to different choice of theory. A theory is fixed once one chooses one of the four 
patterns of BC conditions.

In this paper, we will take the following $(D,D)$ BC for a scalar field,
 \begin{equation}
  \label{DD}
  \delta\Phi
  |_{z=\pm L/2}=0 \;\;\;{\rm and}\;\;\;\Phi
  |_{z=\pm L/2}=v
 \end{equation}
on both branes, where $v$ is a free complex constant of mass dimension $[3/2]$. Notice that these 
BCs are the most general form of Dirichlet BCs which are consistent with the action principle. 
Here, we can determine the solution of EoM under these BCs as 
 \begin{eqnarray}
  \Phi^c(z)=v,
 \end{eqnarray} 
so that the VEV profile on the extra-dimensional direction is flat. How about the profiles of 
quantum modes of the physical Higgs and NG bosons?

\subsection{Profiles of physical Higgs and others}

Let us investigate wave-function profiles of its quantum fluctuation modes that correspond to the 
physical Higgs and NG modes by regarding $\Phi$ as the bulk SM Higgs doublet. The most general form
 of the Dirichlet BC on $\Phi(x,z)$ is
 \begin{align}
  \delta\Phi(x,\pm{L\over2})
	&=0,	&
  \Phi(x,\pm{L\over2})
	&=
   \left(
    \begin{array}{c}
     v_1 \\
     v_2
    \end{array}
   \right),
 \end{align}
where $v_1$ and $v_2$ are free complex constants. Without loss of generality, we can always take a 
basis by an $SU(2)_L\times U(1)_Y$ field redefinition so that the BCs become 
 \begin{align}
  \delta\Phi(x,\pm{L\over2})
	&=0,	\label{DD1} \\
  \Phi(x,\pm{L\over2})
	&=
   \left(
    \begin{array}{c}
     0 \\
     v
    \end{array}
   \right), 
 \end{align}
and thus, the resultant VEV profile is given by
 \begin{eqnarray}
  \Phi^c(z)=
   \left(
    \begin{array}{c}
     0 \\
     v 
    \end{array}
   \right). \label{VEV}
 \end{eqnarray}
The Higgs doublet $\Phi$ is KK-expanded around the VEV in Eq.\eqref{VEV} by infinite number of 4D 
fields as
 \begin{eqnarray}
 \Phi(x,z)=\left(
         \begin{array}{c}
          \displaystyle\sum_{n=0}^\infty f_n^{\varphi^+}(z)\varphi_n^+(x) \\
          v+\frac{1}{\sqrt{2}}\displaystyle
          \sum_{n=0}^\infty[f_n^H(z)H_n(x)+if_n^\chi(z)\chi_n(x)]
         \end{array}
        \right), 
 \end{eqnarray}
where we took $\Phi_R/\sqrt{2}=v$ and $\Phi_I=0$ in the language of the previous subsection, and 
 $H_n(x)$ are the physical Higgs scalar. A linear combination of $\chi_n(x)$ and $Z_z^{(n)}(x)$ and 
that of $\varphi_n^+(x)$ and $W_z^{(n)+}(x)$ are, respectively, absorbed as the longitudinal 
component of $Z_\mu^{(n)}$ and $W_\mu^{(n)+}$, where $Z_z$ and $W_z$ are the fifth components of 5D $Z$
 and $W$ bosons. Notice that the orthogonal combinations to above ones are the physical neutral 
pseudo-scalar and charged bosons, respectively. 

The 5D action for $H(x,z)$ can be written as
 \begin{eqnarray}
  S_H &=& \int d^4x\int_{-L/2}^{+L/2}dz\bigg[\frac{1}{2}H(\Box+\partial_z^2)H 
          +\frac{\delta(z-L/2)}{2}H(-\partial_zH) \nonumber \\
      & & \phantom{\int d^4x\int_{-L/2}^{+L/2}dz\bigg[\frac{1}{2}H(\Box+\partial_z^2)H}
          +\frac{\delta(z+L/2)}{2}H(+\partial_zH)\bigg],
 \end{eqnarray}
and the KK equation for the physical Higgs is given by
 \begin{eqnarray}
  \partial_z^2f_n^H(z)=-\mu_{Hn}^2f_n^H(z),
 \end{eqnarray}
where general solution of this KK equation is
 \begin{eqnarray}
  f_n^H(z)=\alpha_n\cos(\mu_{Hn}z)+\beta_n\sin(\mu_{Hn}z).
 \end{eqnarray}
Taking the Dirichlet BCs in Eq.\eqref{DD1} for quantum fluctuation, we obtain
 \begin{eqnarray}
  f_n^H(z=\pm L/2)=0, \label{DDq}
 \end{eqnarray}
and hence the wave-function profiles become
 \begin{eqnarray}
 f_n^H(z)=\left\{
     \begin{array}{ll}
      \sqrt{\frac{2}{L}}\cos\left(\frac{(n+1)\pi}{L}z\right) &
      \mbox{ for even } n, \\
      \sqrt{\frac{2}{L}}\sin\left(\frac{(n+1)\pi}{L}z\right) &
      \mbox{ for odd } n.
     \end{array}
    \right.\label{fnHz}
 \end{eqnarray}
They are are shown in Table~\ref{tab-pro}.\footnote{\label{comparison_to_previous}They look similar
 to the limit of the large boundary coupling discussed in the setup with brane localized Higgs 
potential \cite{Haba:2009uu}. However, we do not have the large boundary coupling involving the NG 
modes in the current model. The non-flat profile of the physical Higgs field becomes cosine 
function in Eq.\eqref{fnHz}, where the KK number $n$ is shifted by unity from the one in other 
literature, in accordance with~\cite{Haba:2009uu}.} It is worth noting that the wave-function 
profile of the lowest ($n=0$) mode, which corresponds to the SM-like Higgs, is described by a 
cosine function, 
 \begin{eqnarray}
  f_0^H(z)=\sqrt{\frac{2}{L}}\cos\left(\frac{\pi}{L}z\right).  
 \end{eqnarray}
Notice that, if we took Neumann BC for the Higgs, the flat profile of the lowest mode appears.  

\begin{table}[t]
\begin{center}
\begin{tabular}{|c|c|c|c|}
\hline
\multicolumn{4}{|c|}{Typical profiles of Higgs field}\\
\hline
\hline
$n=0$ &$n=1$ & $n=2$ & $n=3$ \\
\includegraphics[scale =0.42]{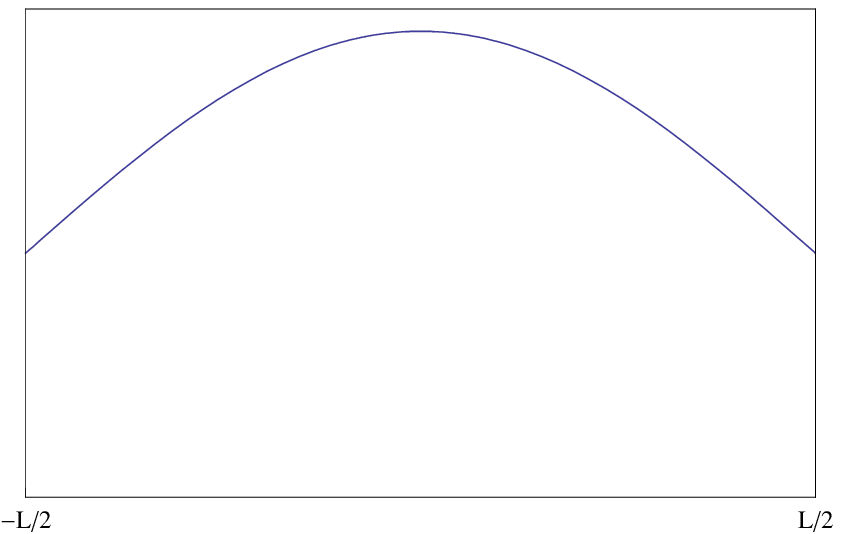} &
\includegraphics[scale =0.42]{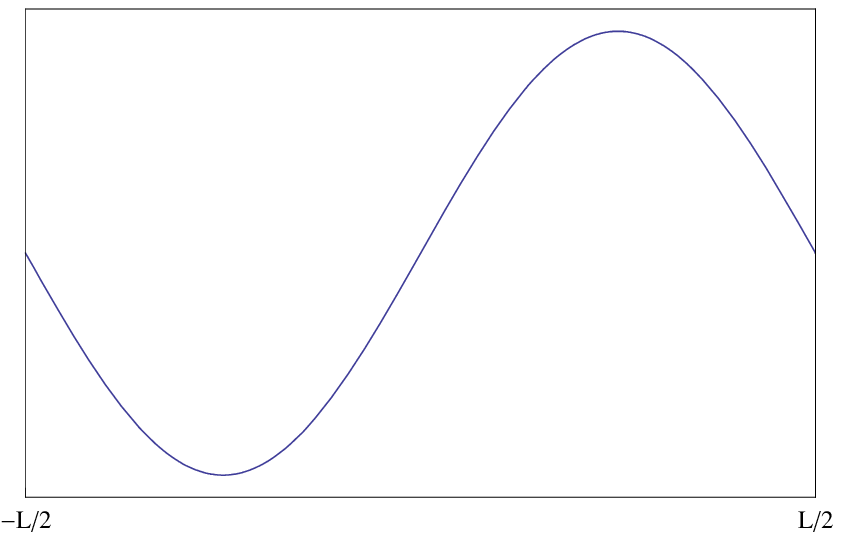} &
\includegraphics[scale =0.42]{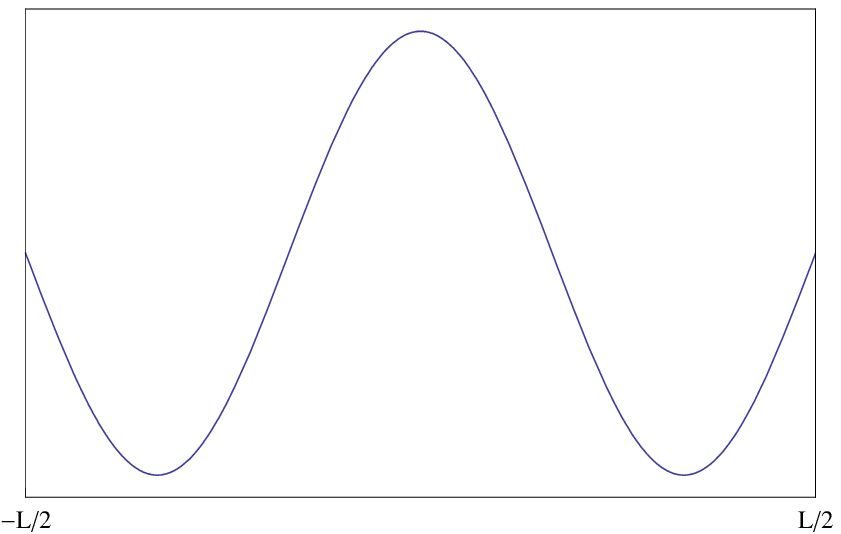} &
\includegraphics[scale =0.42]{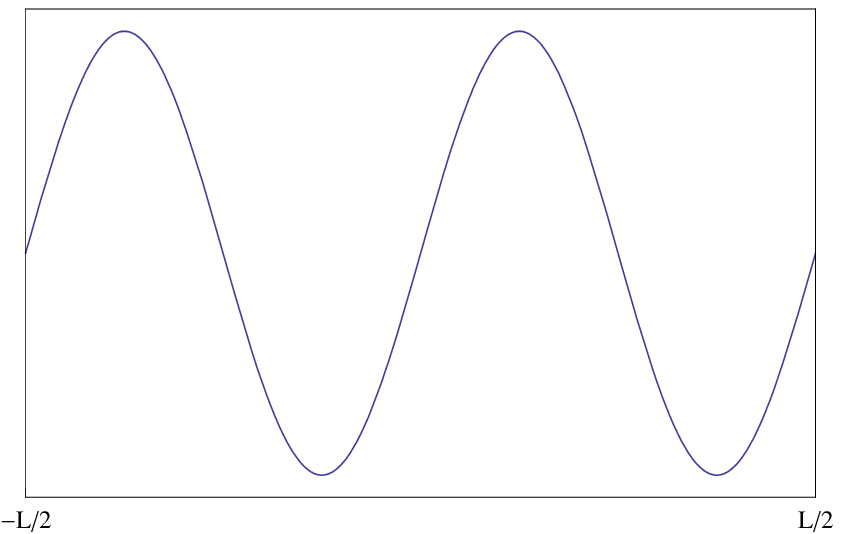} \\
\hline
\end{tabular}
\end{center}
\caption{The wave-function profiles of the physical Higgs}
\label{tab-pro}
\end{table}
This is one of the interesting results from this simple setup of taking Dirichlet BC for the bulk 
Higgs doublet, and this property leads to interesting phenomenology such as the top Yukawa 
deviation.

The $n$-mode Higgs mass is calculated as
 \begin{eqnarray}
  m_{H^{(n)}}^2
  =-\int_{-L/2}^{+L/2}dzf_n^H(z)\partial_z^2f_n^H(z)
  =\left(\frac{(n+1)\pi}{L}\right)^2,
 \end{eqnarray}
where the mass of the lowest ($n=0$) mode Higgs becomes the KK scale:
 \begin{eqnarray}
  m_{KK}\equiv\frac{\pi}{L}.
 \end{eqnarray}
This feature of KK-scale Higgs mass is the specific result induced from the Dirichlet BC in 
Eq.(\ref{DDq}). Note that one does not have a theoretical constraint on the magnitude of the Higgs 
mass from the discussions of perturbative unitarity as in the SM since the mass depends on the 
compactification scale but not on the Higgs self-coupling (see, footnote 
\ref{comparison_to_previous}). 

Profiles of $n$-mode of $\chi^{(n)}$ and $\varphi^+{}^{(n)}$ are the same as $H^{(n)}$, because the 5D 
action of $\chi(x)$ and $\varphi^+(x)$ have the same form as that of $H(x)$. Therefore, 
$f_n^\chi(z)$ and $f_n^{\varphi^+}(z)$ also obey the same KK equations as $f_n^H(z)$. Taking this 
Dirichlet BC is similar to introducing an extra \emph{fake Higgs} field $\phi$ which is localized 
at the boundary and couples to the bulk Higgs doublet as $|\phi|^2\left(|\Phi|^2-v^2\right)$ and 
with a limit of $|\langle \phi \rangle|\rightarrow \infty$. But, a crucially different point is 
that the three- and four-point Higgs self-couplings of $HHH$, $H\chi\chi$, $H\varphi^+\varphi^-$, 
$HHHH$, $HH\chi\chi$, $HH\varphi^+\varphi^-$, $\chi\chi\chi\chi$, $\chi\chi\varphi^+\varphi^-$, and 
$\varphi^+\varphi^-\varphi^+\varphi^-$ vanish in the current model, since there is no Higgs 
potential. (Here and hereafter we represent the lowest KK-modes of the bulk Higgs by $H(x)$, 
$\chi(x)$, and $\varphi^+(x)$, for simplicity.) It means that longitudinal components of gauge 
bosons, $W_L$ and $Z_L$, interact only through the gauge couplings. 

\subsection{Profiles of gauge fields}

Next, we show the gauge sector. We take usual Neumann BC for the gauge fields as
 \begin{equation}
  \label{NN}
  \partial_z A_\mu (z)|_{z=\pm L/2}=0 \;\;\;{\rm and}\;\;\;
  A_z(z)|_{z=\pm L/2}=0.
 \end{equation}
The gauge boson masses are derived from the Higgs kinetic term as 
 \begin{eqnarray}
  S_{\mbox{{\scriptsize kin}}}=\int d^4x\int_{-L/2}^{+L/2}dx|D_M\Phi|^2
 \end{eqnarray}
with the covariant derivative
 \begin{eqnarray}
  D_M\equiv\partial_M+ig_5W_M^aT^a+ig_5'B_MY,
 \end{eqnarray}
where the 5D gauge couplings are related to the 4D ones by $g_5 = g\sqrt{L}$ and 
$g_5' = g'\sqrt{L}$. The SM gauge bosons are given by 
 \begin{eqnarray}
  Z_M\equiv\frac{g_5W_M^3-g_5'B_M}{\sqrt{g_5+g_5'}},~~~
  A_M\equiv\frac{g_5W_M^3+g_5'B_M}{\sqrt{g_5+g_5'}},~~~ 
  W_M^\pm \equiv\frac{W_M^1\mp iW_M^2}{\sqrt{2}}. \label{gauge}
 \end{eqnarray}
Therefore, the KK equation for the weak bosons become 
 \begin{eqnarray}
  \left[\partial_z-\frac{v^2}{2}(g_5^2+g_5'{}^2)\right]f_n^Z(z)
  &=&-\mu_{Zn}^2f_n^Z(z), \\
  \left[\partial_z-\frac{v^2}{2}g_5^2\right]f_n^W(z)&=&-\mu_{Wn}^2f_n^W(z).
 \end{eqnarray}
It is seen that the Neumann BCs for the gauge fields guarantees the flatness of profiles of the 
zero-mode gauge fields. Then, the usual gauge boson masses are given by\footnote{Note that the 
resultant $Z$ and $W$ masses could be correct ones due to the custodial symmetry even under the 
bulk Higgs mass. In this paper, the bulk potential is assumed to be zero, for simplicity.}
 \begin{eqnarray}
  m_Z^2 = \frac{(g^2+g'{}^2)v_{\mbox{{\scriptsize EW}}}^2}{2},~~~
  m_W^2 = \frac{g^2v_{\mbox{{\scriptsize EW}}}^2}{2},
 \end{eqnarray}
where $v=v_{\mbox{\scriptsize EW}}/\sqrt{L}$. The masses of $n$th KK-modes are described as
 \begin{eqnarray}
  m_{Z^{(n)}}=m_Z^2+n^2m_{KK}^2,~~~~~m_{W^{(n)}}=m_W^2+n^2m_{KK}^2, 
 \end{eqnarray}
and we find that the $n$-mode Higgs mass is the same as the $(n+1)$-mode KK gauge boson masses. The
 frequency of $n$-mode Higgs profile is also the same as that of $(n+1)$-mode KK gauge bosons' 
profiles (see footnote~\ref{comparison_to_previous}). These results are originating from the 
Dirichlet (Neumann) BC for the bulk Higgs (gauge) field.

Here we focus on the Higgs mechanism of the zero-mode gauge bosons in more detail. As discussed 
above, although the VEV itself has a flat profile, the 5D fields $\chi(x,z)$ and $\varphi^\pm(x,z)$
 have not zero-mode. Notice also that the lowest mode of the $W_\mu^\pm(x,z)$ and $Z_\mu(x,z)$ are 
flat at the tree level. Therefore, $\varphi$ and $\chi$ must somehow yield the flat profile to be 
absorbed into $W_\mu^{(0)\pm}$ and $Z_\mu^{(0)}$. Key observation is that the NG-boson that is absorbed
 by the zero-mode $W$ (or $Z$) is composed by a linear combination of $n=0$ to $\infty$ modes of NG
 boson. For example, $Z_L^{(0)}$ absorbs the following field having flat profile along the 5D 
direction except at the boundary,
 \begin{eqnarray}
  && \chi (x,z)=0, \hspace{1.5cm}(z=\pm L/2), \label{flat-like1} \\
  && \chi (x,z)=\chi_{\text{NG}}(x), ~~~(-L/2<z<L/2), \label{flat-like2}
 \end{eqnarray}
which can be realized by a superposition of the infinite number of modes
 \begin{eqnarray}
  \chi(x,z)=\sum_{n=0}^{\infty}f_n^\chi(z)\chi_n(x), \label{vz}
 \end{eqnarray}
where
 \begin{eqnarray}
  \chi_n(x)=\frac{2\sqrt{2L}(-1)^{n/2}}{(n+1)\pi}\chi_{\text{NG}}(x),
 \end{eqnarray}
for even $n$.\footnote{$\chi_n(x)$ can be calculated from 
$\chi_n(x)=\int_{-L/2}^{+L/2}dzf_n^\chi(z)\chi_{\text{NG}}(x)$. Thus, there are contributions to 
$\chi_n(x)$ from only even $n$-mode.} By rewriting $n=2m$, we can show 
 \begin{eqnarray}
  \chi(x,z)=\chi_{\text{NG}}(x)\sum_{m=0}^\infty\frac{4(-1)^m}{(2m+1)\pi}\cos
            \left(\frac{(2m+1)\pi}{L}z\right).
 \end{eqnarray}
This profile exactly realizes the flat profile in Eqs.\eqref{flat-like1} and \eqref{flat-like2}. 

\begin{figure}
\hspace{4.1cm}(a)\hspace{7.4cm}(b)
\begin{center}
\includegraphics[scale = 0.8]{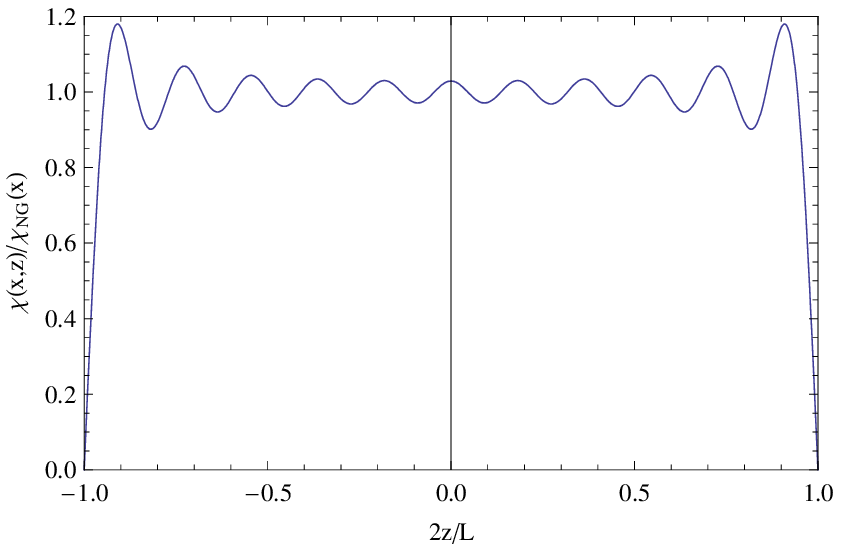}\hspace{1cm}
\includegraphics[scale = 0.8]{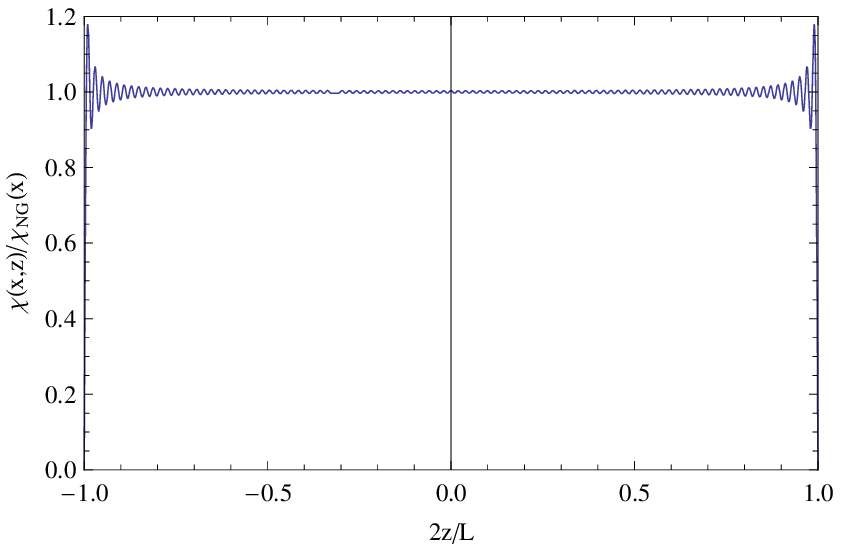}
\end{center}
\hspace{4.1cm}(c)\hspace{7.4cm}(d)
\begin{center}
\includegraphics[scale = 0.8]{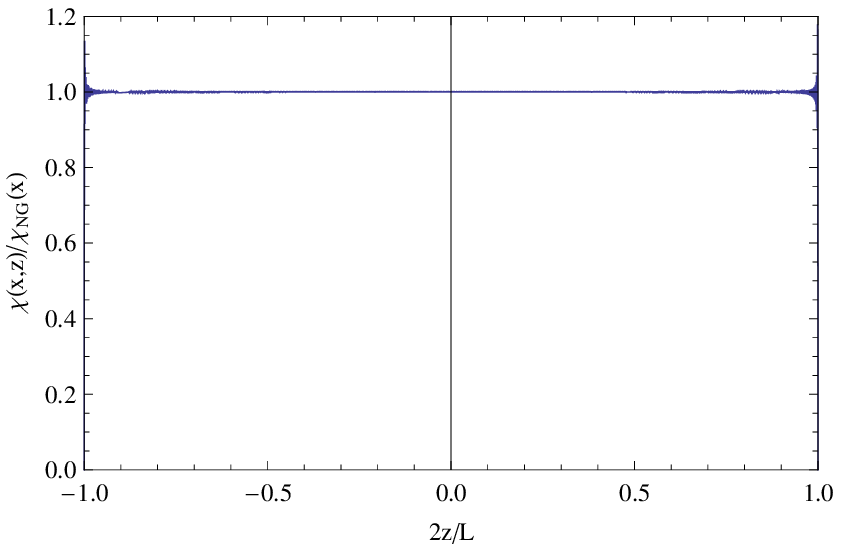}\hspace{1cm}
\includegraphics[scale = 0.8]{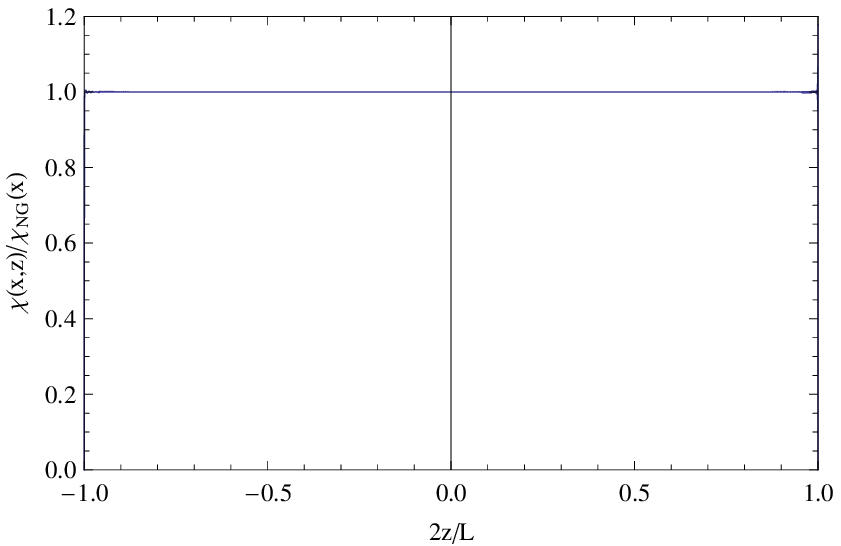}
\end{center}
\caption{The relation between the flatness of VEV profile and largeness of $m$. 
The horizontal and vertical axes are $2z/L$ and $\chi(x,z)/\chi_{\text{NG}}(x)$, respectively, in all
 figures. The figure (a) corresponds to the case of $m=10$, (b) for $m=100$, (c) for $m=1000$, and 
(d) for $m=10000$.}
\label{fig1-1}
\end{figure}

The 5D gauge theory is non-renormalizable and hence is defined with a ultraviolet cutoff. 
Therefore, one might think that the sum should be taken up to a finite value of $m$ corresponding 
to a cutoff scale $\Lambda=mm_{KK}=m\pi/L$. Let us see the correspondence between the cutoff scale 
$\Lambda$, equivalently the maximum $m$, and the flatness of the would-be NG mode 
$\chi(x,z)$,\footnote{An $n$-th partial sum of the Fourier series has an oscillation near the 
discontinuity points which is known as the Gibbs phenomenon. This is the reason why we must 
consider higher KK-tower to realize the profile of Eqs.(\ref{flat-like1}) and (\ref{flat-like2}).} 
which is represented in Fig.~\ref{fig1-1}. 
In our setup, the KK scale is a model parameter constrained from the precision EW measurements. For
 example, when $m=10^4$ and $m_{KK}=600$ ($4\times10^3$) GeV are taken, the cutoff scale of this 
setup becomes $\Lambda=6\times10^3$ ($4\times10^4$) TeV. 

Then, what should we think the 5D cutoff? There must exist heavier KK-modes than the cutoff scale 
in principle, otherwise the complete flat profile of NG bosons could not be obtained. There is 
generally a high energy cutoff in this 5D model so that we cannot sum over infinite number of 
KK-modes. However, a model with a cutoff $\Lambda$ is expected to have an ambiguity of 
${\mathcal O}(\Lambda^{-1})$, which are induced from integrating out heavier KK-modes above 
$\Lambda$. In order to distinguish this ambiguity, for example, deviation from flat profile in 
above case, we need an experimental resolution of ${\mathcal O}(\Lambda^{-1})$, so that the 
ambiguity could be neglected.

As for longitudinal components of $W_L^{(n)}$ and $ Z_L^{(n)}$ ($n\geq 1$), they are mainly composed 
by $W_y^{(n)}$ and $Z_y^{(n)}$, respectively. The orthogonal linear combinations are physical charged 
KK scalar and neutral pseudo-scalar particles, respectively. 

In our model, the gauge symmetry is violated by the extra-dimensional BC of Higgs. However, the 5D 
 gauge symmetry will be restored as an energy scale becomes much higher than the KK scale. We note 
that in several models of orbifold/boundary symmetry breaking, it has been shown that the 
longitudinal gauge boson scattering etc. are indeed unitarized by taking into account KK-mode 
contributions \cite{Hall}-\cite{2009ei}. Therefore, it would be expected that the bulk gauge boson 
scattering is unitarized in such a region of our model (above the KK scale but lower than the 5D 
cut-off scale) by taking account all the relevant KK-modes.

At the end of this section, we comment on the KK parity. It is known that the 
UED model conserves KK parity. How about our setup? In our model, the Dirichlet
 BC is taken for the Higgs field, and when we take the same BC on both $z=\pm 
L/2$ branes, there exists a reflection symmetry between the branes. This 
guarantees the conservation of the KK parity in the gauge and Higgs sector. 
Thus, the existence of KK parity in the Lagrangian depends on a fermion sector.
 When the fermions are localized on the 4D branes, the KK parity is broken in 
general. On the other hand, the KK parity is conserved in a bulk fermion setup 
similar to the UED model. When KK parity exists in the model, the lightest KK 
parity odd particle is stable, which can be a candidate of the dark matter. We 
will return to this point later.

\section{Phenomenological aspects}

In this section, we discuss phenomenological aspects of above extra-dimensional setup, namely top 
Yukawa deviation, the production and decay of the Higgs, constraints from the EW precision 
measurements, and dark matter candidate.

\subsection{Top Yukawa deviation}

Let us investigate the top Yukawa deviation, which is one of the most interesting phenomenologies 
in our setup. First we give a brief explanation about the top Yukawa deviation. In the SM, 
the fermion mass terms and couplings with physical Higgs are given by
 \begin{eqnarray}
  -\mathcal{L}_Y^{\mbox{{\scriptsize SM}}}
   \supset m_f\bar{f}(x)f(x)+y_f\frac{H(x)}{\sqrt{2}}\bar{f}(x)f(x),
 \end{eqnarray}
where $m_f$ is the mass of a SM fermion $f(x)$. The mass of fermion is described by the VEV of the 
Higgs multiplied by the Yukawa coupling, $m_f\equiv y_fv_{\text{EW}}$. This Yukawa coupling also 
determines the magnitude of the coupling between fermion and the physical Higgs field, 
 $y_{\bar{f}fH}\frac{H(x)}{\sqrt{2}}\bar{f}f$, and, thus
 \begin{align}
  y_{\bar{f}fH}=y_f. 
  \label{SM_expectation}
 \end{align}
This is the simple tree level result in the SM. Among the Yukawa couplings in 
the SM, that of top quark is the largest. Therefore, this SM expectation in 
Eq.\eqref{SM_expectation} should be checked by precisely observing the Higgs 
events related with the top quarks at the LHC. The deviation of the expectation
 from the SM is called top Yukawa deviation and observation of this phenomena 
would suggest the presence of physics beyond the SM. Here, we focus on the top 
quark when we discuss the Yukawa deviation.

It is well known that such a deviation can generally occur in multi-Higgs model. A popular 
multi-Higgs model is the Minimal Supersymmetric Standard Model (MSSM), in which the coupling of the
 top quark with the physical Higgs field is described as 
 \begin{eqnarray}
  -\mathcal{L}_t^{\mbox{{\scriptsize MSSM}}}
  \supset y_t\cos\alpha\frac{h^0}{\sqrt{2}}\bar{t}t,
 \end{eqnarray}
while the Yukawa coupling of top quark is determined by 
$y_t=m_t/v_u$. It is seen that the additional factor, $\cos\alpha$, 
is multiplied to the Yukawa coupling. The physical meaning of this 
factor is a mixing angle between the weak eigenstate of the Higgs 
$(H_u^0,H_d^0)$ and the mass eigenstate $(h^0,H^0)$ where $h^0$ is 
the lightest CP-even Higgs field. Such a situation generally occurs 
in other multi-Higgs doublet models, but never happens in 4D 
one-Higgs doublet model.

In our model described with the Dirichlet BC, the Yukawa interaction for the 
top quark and the Higgs field are written as
 \begin{eqnarray}
  -\mathcal{L}_t=y_{t,5}\int_{-L/2}^{+L/2}dz
                 \left[v+f_0^H(z)\frac{H(x)}{\sqrt{2}}\right]
                 \bar{t}(x,z)t(x,z).
 \end{eqnarray}
The top quark mass, $m_t$, and effective top coupling in 4D, $y_t$, can be 
obtained as $m_t=y_{t,5}v$ and 
$y_t=\frac{m_t}{v\sqrt{L}}=\frac{y_{t,5}}{\sqrt{L}}$. Then the ratio of the top
 Yukawa coupling in our model to that of the SM, $r_{H\bar{t}t}$, is given by
 \begin{eqnarray}
  \label{rhtt}
  r_{H\bar{t}t} 
  &=& \frac{1}{\sqrt{L}}\int_{-L/2}^{L/2}dzf_0^H(z)
      =  \frac{2\sqrt{2}}{\pi}\simeq 0.90,
 \end{eqnarray}
which means the top deviation in this setup is 10\% decrease from the 
SM.\footnote{If the SM fermions are brane localized fields, the coupling 
between the top quark and Higgs boson in 4D is completely vanishing 
\cite{Haba:2009pb}. Such a kind of model can predict a maximal top Yukawa 
deviation, $r_{H\bar{t}t}=0$.} Such result can be hardly happened in other 
{\it one-Higgs doublet} models.\footnote{A specific warped 
gauge-Higgs unification model also induces the maximal top Yukawa deviation 
\cite{Hosotani:2008by}. However, the physical results are completely different,
 for example, the Higgs in \cite{Hosotani:2008by} is stable, while the one in 
our model decays quite rapidly as shown later.}

\subsection{Higgs production and decay}

Here we consider the SM-like Higgs production, higher KK Higgs production, and Higgs decay at the 
LHC experiment. At first, let us show the Higgs production processes. 

\subsubsection{Higgs production}

\begin{figure}[t]
\hspace{3.7cm}(a) gluon fusion\hspace{2cm}(b) vector boson fusion
\begin{center}
\includegraphics[scale = 0.8]{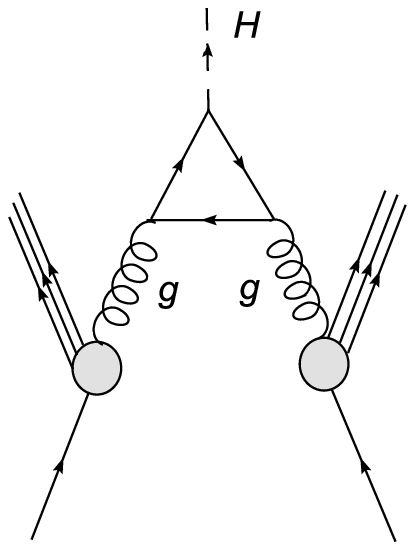}\hspace{2cm}
\includegraphics[scale = 0.8]{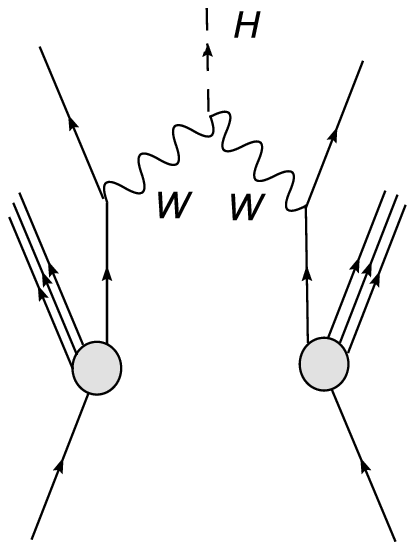}
\end{center}
\caption{Dominant processes of Higgs production at the LHC}
\label{fig7}
\end{figure}

The Higgs boson can be produced mainly through the following four different channels
 \begin{itemize}
  \item gluon fusion $gg\rightarrow H$ shown in Fig.~\ref{fig7} (a),
  \item vector boson fusion $qq\rightarrow qqH$ shown in Fig.~\ref{fig7} (b),
  \item associated production with the weak gauge bosons $q\bar{q}\rightarrow WH,~ZH$,
  \item associated production with the heavy quarks $gg,~q\bar{q}\rightarrow 
        t\bar{t}H,~b\bar{b}H$ and $gb\rightarrow bH$.
 \end{itemize}
The SM predicts that the gluon fusion with the top quark 1-loop diagram dominates up to 1 TeV of 
Higgs mass by taking account of leading and next-to-leading order of QCD corrections with 
 $\sqrt{s}=14$ TeV and $m_t=174$ GeV, while the vector boson fusion is a subdominant process for 
the Higgs production. So how is going in our extra-dimension scenario? Since the Yukawa couplings
 of Higgs with  the top quark are modified, the processes for the Higgs production must be 
reanalyzed.

The Higgs production cross section through the gluon fusion depends on the 
magnitude of the coupling between the top quarks and physical Higgs boson. 
Since the coupling between the top quarks and Higgs boson deviates from the top
 Yukawa coupling as is pointed above, the cross section of the gluon fusion 
process becomes 81\% of the SM due to $r_{H\bar{t}t}\simeq0.81$. This is one of
 the most important predictions of our model. As regarding the vector boson 
fusion process, the magnitude of cross section changes because the coupling 
between $W$ and Higgs is also modified. The interaction is written by
 \begin{eqnarray}
  -\mathcal{L}_{WWH}
   =\frac{em_W}{2\sin\theta_W}\frac{1}{2L}\int_{-L/2}^{+L/2}dz
    f_0^H(z)f_0^{W^+}f_0^{W^-}H(x)W^+(x)W^-(x)+h.c.
 \end{eqnarray}
When we write the ratio of the $WWH$ coupling in the 5D model to the SM as 
$r_{WWH}$, it is given by
 \begin{eqnarray}
  r_{WWH}\equiv\frac{1}{\sqrt{L}}\int_{-L/2}^{+L/2}dzf_0^H(z),
 \end{eqnarray}
where we used the fact that the zero-mode profile of $W$ boson is flat 
$f_0^{W^\pm}(z)=1/\sqrt{L}$. Notice that this ratio is the same as 
$r_{H\bar{t}t}$ in Eq.(\ref{rhtt}) and its value is $\simeq0.90$. Therefore, 
the cross section for the Higgs production in the model are overall decreased 
to $81\%$ of the SM, while the branching ratios are not changed. It is because 
other possible processes for the production such as $q\bar{q}\rightarrow HW$, 
$q\bar{q}\rightarrow HZ$, $gg,~q\bar{q}\rightarrow t\bar{t}H,~b\bar{b}H$, and 
$gb\rightarrow bH$ are also suppressed in the same manner.

\subsubsection{Higher KK Higgs production}

Let us consider the Higgs production of higher KK-modes, which might be 
discovered at the LHC and/or
 future international linear collider (ILC) experiment.
In above discussions, we only consider the lightest states of Higgs and top 
 quark propagate in the loop of the gluon fusion process. 
However, in general,
 higher KK Higgs and heavy quarks can contribute in this process. 

In our model, a higher KK Higgs can be produced through the gluon fusion as
 \begin{eqnarray}
  &&q^{(2n)}\bar{q}^{(2|n-l-1|)}    \rightarrow H^{(2(l+1))}, \\
  &&q^{(2n)}\bar{q}^{(2|n-l|+1)}    \rightarrow H^{(2l+1)}, \\ 
  &&q^{(2n+1)}\bar{q}^{(2|n-l|)}    \rightarrow H^{(2l+1)}, \\
  &&q^{(2n+1)}\bar{q}^{(|2(n-l)-1|)} \rightarrow H^{(2(l+1))},
 \end{eqnarray} 
where $q$ are the heavy quarks circulating in the loop shown in Fig. \ref{fig7}
 (a), and $l$, $m$, and $n$ are $0,1,2,\cdots$, which indicate the KK 
number.\footnote{We are discussing in a basis where mixings among KK states are
 allowed in interactions of KK-quarks with KK-gluon.} The cross section 
(coupling) for each process is also suppressed by effect from the deformed 
Higgs profile compared with the one in the UED case. For the vector boson 
fusion, the following KK parity conserving processes are possible,
 \begin{eqnarray}
  && W^{(2n)}W^{(2|n-l-1|)}\rightarrow H^{(2(l+1))}, \\
  && W^{(2n)}W^{(2|n-l|+1)}\rightarrow H^{(2l+1)}, \\
  && W^{(2n+1)}W^{(2|n-l|+1)}\rightarrow H^{(2l+1)}, \\
  && W^{(2n+1)}W^{(|2(n-l)-1|)}\rightarrow H^{(2(l+1))}.
 \end{eqnarray}



\begin{figure}
\hspace{5.4cm}$e^{-(0)}$\hspace{3.9cm}$Z^{(l)}$\vspace{5mm}

\hspace{7.8cm}$Z^{(n)}$\vspace{-1.2cm}

\begin{center}
\includegraphics[scale = 1]{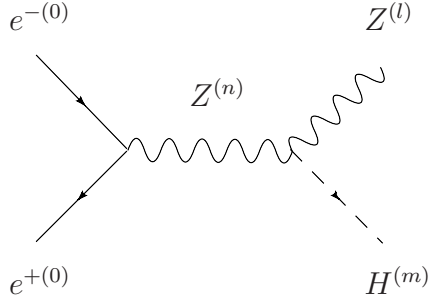}
\end{center}\vspace{-3mm}

\hspace{5.4cm}$e^{+(0)}$\hspace{3.9cm}$H^{(m)}$

\caption{Associated production of KK Higgs}
\label{associated}
\end{figure}

Finally, we comment on the associated production of Higgs boson including 
higher KK-mode shown in  Fig. \ref{associated}. It may be expected that the 
Higgs boson are produced through associated process of 
 \begin{eqnarray}
  e^{+(0)}e^{-(0)}\rightarrow Z^{(n)*} \rightarrow Z^{(l)} H^{(m)}
 \end{eqnarray}
in the future linear collier experiment.
Taking account of KK number, the internal $Z^{(n)}$ must be $Z^{(0)}$ in the.
The final $Z^{(l)}$ and $H^{(m)}$ must be 
 the same KK-mode ($l=m$).

\subsubsection{Higgs decay}

Next, let us study the Higgs decay. In the SM, the dominant decay process of Higgs boson is 
$H\rightarrow b\bar{b}$ for relatively small Higgs mass as $m_H\lesssim140$ GeV. The both branching
 ratios for $H\rightarrow\tau^+\tau^-$ and $H\rightarrow gg$ can reach to about $0.08$ at $100$ 
GeV$\lesssim m_H\lesssim120$ GeV. The Higgs decays into two photons give clean experimental 
signatures whose branching fraction becomes $2\times10^{-3}$ at light Higgs boson masses. For a 
relatively heavy Higgs mass as $350$ GeV$\lesssim m_H$, the first, second, and third dominant 
processes are $H\rightarrow WW$, $ZZ$, and $t\bar{t}$, respectively. The branching fractions of 
these processes must be compared with our case since the Higgs mass of our model is the same as the
 KK scale that is larger than the weak scale. The decay widths of the Higgs to the $WW$ and $ZZ$ 
pair is given by 
 \begin{eqnarray}
  \Gamma_{H\rightarrow W_TW_T}
   &=& \frac{G_Fm_W^4}{2\sqrt{2}\pi m_H}\left(1-\frac{4m_W^2}{m_H^2}\right)^{1/2},
       \label{transverse}\\
  \Gamma_{H\rightarrow W_LW_L}
   &=& \frac{G_Fm_H^3}{8\sqrt{2}\pi}\left(1-\frac{2m_W^2}{m_H^2}\right)^2
       \left(1-\frac{4m_W^2}{m_H^2}\right)^{1/2}, \\
  \Gamma_{H\rightarrow WW}
   &=& 2\Gamma(H\rightarrow W_TW_T)+\Gamma(H\rightarrow W_LW_L) \nonumber \\
   &=& \frac{G_Fm_H^3}{8\sqrt{2}\pi}
       \left(1-\frac{4m_W^2}{m_H^2}+\frac{12m_W^4}{m_H^4}\right)
       \left(1-\frac{4m_W^2}{m_H^2}\right)^{1/2}, 
 \end{eqnarray}
and similarly
 \begin{eqnarray}
  \Gamma_{H\rightarrow ZZ}
   &=& \frac{G_Fm_H^3}{16\sqrt{2}\pi}
       \left(1-\frac{4m_Z^2}{m_H^2}+\frac{12m_Z^4}{m_H^4}\right)
       \left(1-\frac{4m_Z^2}{m_H^2}\right)^{1/2},
 \end{eqnarray}
in the SM, where $W_T$ and $W_L$ mean the transverse and longitudinal modes, 
respectively.

The decay channel $H\rightarrow ZZ\rightarrow 4l$ is important for an actual 
Higgs boson discovery at the LHC experiment.\footnote{For the Higgs mass region
 of $115$ GeV$\lesssim m_H\lesssim2m_Z$, the decay channel can give a clean 
signature. Assuming a magnitude of integrated luminosity as $30\mbox{ 
fb}^{-1}$, the signal can be detected with a significance of more than 
$5\sigma$ for $130$ GeV$\lesssim m_H\lesssim180$ GeV, except in a narrow range 
around $170\,\text{GeV}$ where the decay mode into $WW$ opens.} For the Higgs 
mass region of $180$ GeV$\lesssim m_H\lesssim700$ GeV, the decay mode into 
$4l$ is a reliable one for the discovery of SM Higgs boson at the LHC 
experiment. For a larger Higgs mass, the decay width becomes very large but the 
Higgs might be discovered through other decay modes such as $H\rightarrow 
ZZ\rightarrow ll\nu\nu$ and $H\rightarrow ZZ\rightarrow lljj$ 
\cite{atlas,Ball:2007zza}. As we will discuss later, since the Higgs mass of our
 model should be $430\mbox{ GeV}\lesssim m_H\lesssim 500\mbox{ GeV}$, the 
reliability of SM Higgs boson discovery should be compared in this range.

In our model, we have the overall suppression factor for $H\rightarrow WW$, 
$ZZ$, and $t\bar{t}$ processes. Contrary to the SM, the $H\rightarrow WW(ZZ)$ 
process of our model occurs only through gauge interactions, since the couplings
 of NG bosons such as $H\varphi^+\varphi^-(H\chi\chi)$ vanish. The decay width 
is estimated as
 \begin{equation}
  \Gamma_{H\rightarrow W^+W^-}\simeq\frac{G_Fm_H^3}{8\sqrt{2}\pi}r_{WWH}^2.
  \label{width}
 \end{equation}
Since the Higgs has the same scale mass as the KK gauge bosons in this setup, 
the Higgs mass must be enough large to be consistent with the electroweak 
precision measurements. The mass should be $430\mbox{ GeV}\lesssim m_H\lesssim 
500\mbox{ GeV}$. The width Eq.\eqref{width} is smaller than that of SM because 
of the factor $r_{WWH}^2$.

Reminding
 that $m_H=m_{KK}\gg m_W$ in our model,
 one might think that the Higgs 
 decay process was equivalent to the process of  
 $H\rightarrow\varphi_{\text{NG}}^+\varphi_{\text{NG}}^-$ 
 based 
 on the equivalence theorem for the NG boson,
 where 
 $\varphi_{\text{NG}}^\pm$ is the NG mode eaten by the
 lowest mode of $W^\pm$. 
Since there exists no Higgs 
potential in our model at the tree level, the Higgs cannot couple to 
$\varphi_{\text{NG}}^+\varphi_{\text{NG}}^-$, which might be interpreted that the 
Higgs would decay to two gauge bosons only through the transverse mode of gauge 
bosons and that
 the Higgs decay width was expressed as~Eq.\eqref{transverse}. 
It is, 
 however, not accurate. 
The would-be NG bosons $\varphi_{\text{NG}}^\pm$ are 
 absorbed into $W^\pm$ whose wave-function profiles are given by 
 Eqs.\eqref{flat-like1} and \eqref{flat-like2}. 
As shown in the previous section,
 it is realized by the linear combination of all higher KK-modes, which 
means that a lot of heavier KK states other than the lowest mode are included.
The equivalence theorem does not hold for the decay of the lowest physical Higgs mode.
The Higgs can decay into $W^+W^-$ through the longitudinal component
 of $W^\pm$, whose decay width dominates over the one through the transverse 
modes. The total decay width can be well approximated as Eq.\eqref{width}.

%


\subsection{Electroweak precision measurements}

\begin{figure}
\begin{center}
\includegraphics[scale = 1]{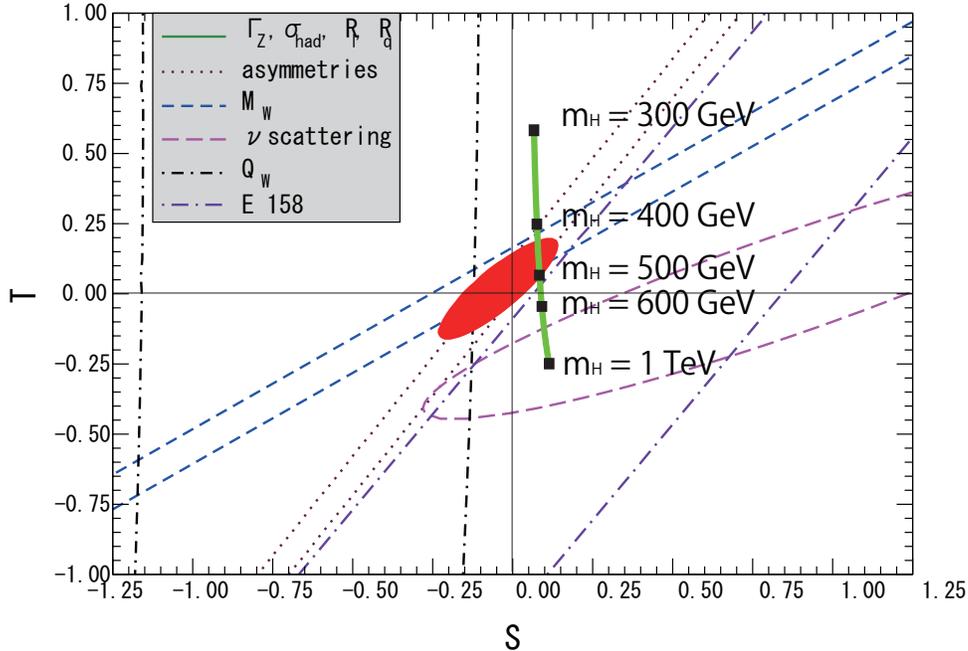}
\end{center}
\caption{$S$ and $T$ plot: Contours show $1\sigma$ constraints 
(39.35 \%) from various inputs except for the central one 
representing 90\% CL allowed by all data~\cite{pdg}.}
\label{fig1}
\end{figure}

Next, let us discuss constraints from EW precision measurements on this setup. 
Here, we estimate the $S$ and $T$ parameters \cite{stu2,stu3,stu1} in
 the model, which are defined as
 \begin{eqnarray}
  \alpha S &\equiv& 4e^2[\Pi_{33}^{\mbox{{\scriptsize new}}}{}'(0)
                    -\Pi_{3Q}^{\mbox{{\scriptsize new}}}{}'(0)], \\
  \alpha T &\equiv& \frac{e^2}{s^2c^2m_Z^2}
                    [\Pi_{11}^{\mbox{{\scriptsize new}}}(0)
                    -\Pi_{33}^{\mbox{{\scriptsize new}}}(0)],
 \end{eqnarray}
where
 \begin{eqnarray}
  s\equiv\sin\theta_W\equiv\frac{g'}{\sqrt{g^2+g'{}^2}},~~~
  c\equiv\cos\theta_W\equiv\frac{g}{\sqrt{g^2+g'{}^2}}.
 \end{eqnarray}
$\Pi_{XY}(q^2)$ is the vacuum polarizations, which can be described as
 \begin{eqnarray}
 \Pi_{XY}(q^2)\equiv\Pi_{XY}(0)+q^2\Pi_{XY}'(q^2),
 \end{eqnarray}
and $\Pi_{XY}'(q^2)$ is equal to $d\Pi_{XY}/dq^2$ at $q^2=0$. The 
$\Pi_{11}$ and $\Pi_{33}$ are represented by
 \begin{eqnarray}
  \Pi_{11}=\frac{s^3}{e^2}\Pi_{WW} ~~~\mbox{ and }~~~
  \Pi_{33}=\frac{s^3}{e^2}[c^2\Pi_{ZZ}+2sc\Pi_{ZA}+s^2\Pi_{AA}],
 \end{eqnarray}
respectively. In our setup, $S$ and $T$ parameters are roughly 
estimated as \cite{Gogoladze:2006br,Haba:2009uu}
\begin{eqnarray}
  S &\simeq& \frac{1}{6\pi}\log\left(\frac{m_H}{m_{H,\text{ref}}}\right)+\sum_{n=1}^\infty\frac{1}{4\pi}f_S^{\text{KK-top}}\left(\frac{m_t^2}{n^2m_{KK}^2}\right), \\
  T &\simeq& -\frac{3}{8\pi c^2}\log\left(\frac{m_H}{m_{H,\text{ref}}}\right)+\sum_{n=1}^\infty\frac{3m_t^2}{16\pi^2v_{\text{EW}}^2}\frac{1}{\alpha}f_T^{\text{KK-top}}\left(\frac{m_t^2}{n^2m_{KK}^2}\right),  
 \end{eqnarray}
where $m_H=m_{KK}$. $m_{H,\text{ref}}$ is the reference Higgs mass taken as 
$m_{H,\text{ref}}=117$ GeV, and
 \begin{eqnarray}
  f_S^{\text{KK-top}}(z) &=& \frac{2z}{1+z}-\frac{4}{3}\log(1+z), \\
  f_T^{\text{KK-top}}(z) &=& 1-\frac{2}{z}+\frac{2}{z^2}\log(1+z).
 \end{eqnarray}
The first terms in both S and T parameters approximate the 
absence of the SM Higgs, replacing it by the first KK Higgs $H^{(0)}$ whose 
coupling to SM zero modes is smaller by a factor $0.9$, and the second terms 
are the KK top ones. Since a contribution to $S$ and $T$ parameter from the KK 
Higgs modes are small at $m_{\text{KK}}\lesssim500$ GeV, we drop the corresponding
 terms. We have not truncated the KK sum but performed it for infinite modes. 
Generically this is known to be a good strategy that does not spoil the five 
dimensional gauge symmetry at short distances. Notice that Higgs mass and KK top
 contributions are negative and positive for $T$ parameter, respectively, while 
they contribute positive for $S$ parameter. This is the reason why a heavy Higgs
 is consistent with $S$ and $T$ parameters contrary to the SM. The $(S,T)$ plot 
in this setup is presented in Fig. \ref{fig1}.
We plot the parameters in a region of $300\,\text{GeV}\leq m_H 
\leq 1\,\text{TeV}$, and take the reference Higgs mass as 117 GeV. We find that 
the first KK Higgs mass $m_H=m_{KK}$ is constrained in the range 
$430\,\text{GeV}\lesssim m_H\lesssim500\,\text{GeV}$ within 90\% CL. Our plot 
roughly corresponds a diagonal line $m_H=1/R$ in Fig.~3 of 
Ref.~\cite{Appelquist}, which gives 
$370\,\text{GeV}\lesssim m_H=1/R\lesssim550\,\text{GeV}$ within 90\% CL from the
 data 2002.

\subsection{Dark matter}
We comment on a dark matter (DM) candidate in our model. In the UED 
model, the lightest KK particle (LKP) can be a candidate for the DM 
because the KK parity is always conserved. The possibility that the 
KK photon becomes the LKP, and thus the DM candidate, has been widely
 discussed in a lot of literatures (see e.g.~\cite{Hooper:2007qk} for
 review and references).

If there is a reflection symmetry for the BCs in this scenario, the KK parity is
 conserved, and thus the LKP with odd ($n=1$) parity is stable, which can be a 
DM candidate. The masses of $n$th KK excitation states, $X^{(n)}$, of SM field 
are given by
 \begin{eqnarray}
  m_{X^{(n)}}^2=n^2m_{KK}^2+m_{X^{(0)}}^2, \label{rc}
 \end{eqnarray}
at tree level, where $X^{(0)}$ and $m_{X^{(0)}}$ indicate the 
corresponding SM field and its mass. The mass of the neutral Higgs 
can be rewritten as
 \begin{eqnarray}
  m_{H^{(n)}}^2=m_{H^{(n)}}^2+\delta m_{H^{(n)}}^2, \label{KKH}
 \end{eqnarray}
including the radiative corrections to the KK masses induced from 
loop diagrams traversing around the extra dimension. 
On the other hand, the correction to the KK photon is given by 
\cite{Cheng:2002iz}
 \begin{eqnarray}
  \delta m_{A^{(n)}}^2 
   &=& \frac{1}{2}\Bigg[\delta m_{B^{(n)}}^2+\delta m_{W^{(n)}}^2
       +\frac{1}{2}(g^2+g'{}^2)v_{\text{EW}}^2 \nonumber \\
   & & \phantom{\frac{1}{2}\Bigg[}
       +\sqrt{\left\{\delta m_{B^{(n)}}^2-\delta m_{W^{(n)}}^2
       +\frac{1}{2}(g^2+g'{}^2)v_{\text{EW}}^2
        \right\}^2-(2gg'v_{\text{EW}}^2)^2}\Bigg],
 \end{eqnarray}
where
 \begin{eqnarray}
  \delta m_{B^{(n)}}^2 
   &=& -\frac{39}{2}\frac{g'{}^2\zeta(3)}{16\pi^4}m_{KK}^2
       -\frac{1}{3}\frac{g'{}^2}{16\pi^2}n^2m_{KK}^2\ln
       \frac{\Lambda}{nm_{KK}}, \\
  \delta m_{W^{(n)}}^2 
   &=& -\frac{5}{2}\frac{g^2\zeta(3)}{16\pi^4}m_{KK}^2
       +15\frac{g^2}{16\pi^2}n^2m_{KK}^2\ln\frac{\Lambda}{nm_{KK}}.
 \end{eqnarray}
It is found that the LKP in our scenario is the KK photon naively since the mass
 of KK photon is similar to the UED case and further the Higgs 
mass does not aquire negative contribution from the quartic coupling as in UED.

\subsection{FCNC problem}

Finally, let us comment that our setup does not suffer from serious 
flavor changing neutral currents (FCNCs) problem since the model 
have only one Higgs doublet. We note that in our  
scenario, all the bulk fermions have a flat wave-function profile at 
the tree level where the bulk and brane Higgs potentials are not 
generated via renormalization group running. Therefore there are no 
tree level FCNC processes coming from the overlap integral along the 
extra-dimension. This is not affected by the fact that Higgs VEV and 
physical field have different profile. One might think that 
contribution from a loop of charged KK Higgs field is different from 
that in the UED model. However, the only difference is coming from 
the change of the BC from Neumann to Dirichlet. Therefore, the 
overlap integral of two $n$th-modes and a single (flat) zero-mode are
 identical in these two models, giving the same coupling. FCNC loop 
contributions from KK-modes are not different from the ordinary UED 
model \cite{Buras:2003wc}-\cite{Colangelo:2007jy}.

\subsection{Comparison of models}

At the end of section, we tabulate phenomenological results of our 
model in comparison with
other extra-dimensional models \cite{Haba:2009uu}. The models proposed in 
ref \cite{Haba:2009uu} have been considered in flat 5D spacetime with 
brane localized Higgs potential (BLP). The models of BLP
have been classified into the BLF case and bulk fermion (BF) case, the latter 
corresponding to a more general setup of the UED model. We refer to the models 
given in \cite{Haba:2009uu} as BLP(BLF) and BLP(BF) in this subsection. The 
phenomenological aspects of our model and comparisons of it with BLP(BLF) and 
BLP(BF) scenarios are given in Table \ref{tab3}.

\begin{table}
\begin{center}
\begin{tabular}{|l||c|c|}
\hline
 & BLP(BLF) & BLP(BF) \\
\hline
\hline
Top Yukawa & Tiny in small $\hat{\lambda}$ & Tiny in $\hat{\lambda}$ \\
deviation & $90\%$ in large $\hat{\lambda}$ & $8\%$ in large $\hat{\lambda}$ \\
\hline
$H^{(0)}$ production & Same as SM in small $\hat{\lambda}$ & Same as SM in small $\hat{\lambda}$ \\
by $gg$ fusion & $1\%$ of SM in large $\hat{\lambda}$ & $85\%$ of SM in large $\hat{\lambda}$  \\
\hline
$H^{(0)}$ production & Same as SM in small $\hat{\lambda}$ & Same as SM in small $\hat{\lambda}$ \\
by $WW$ fusion & $85\%$ of SM in large $\hat{\lambda}$ & $85\%$ of SM in large $\hat{\lambda}$ \\
\hline
$H^{(n)}$ production &  & $q^{(2n)}\bar{q}^{(2m)}\rightarrow H^{(2(l+1))}$ \\
by $gg$ fusion & $q^{(n+1)}\bar{q}^{(m+1)}\rightarrow H^{(l+1)}$ in small $\hat{\lambda}$ & $q^{(2n)}\bar{q}^{(2m+1)}\rightarrow H^{(2l+1)}$\\
 & & $q^{(2n+1)}\bar{q}^{(2m+1)}\rightarrow H^{(2(l+1))}$\\
\hline 
$H^{(n)}$ production & $W^{(2n)}W^{(2m)}\rightarrow H^{(2(l+1))}$ & $W^{(2n)}W^{(2m)}\rightarrow H^{(2(l+1))}$ \\
by $WW$ fusion & $W^{(2n)}W^{(2m+1)}\rightarrow H^{(2l+1)}$ & $W^{(2n)}W^{(2m+1)}\rightarrow H^{(2l+1)}$ \\
 & $W^{(2n+1)}W^{(2m+1)}\rightarrow H^{(2(l+1))}$ & $W^{(2n+1)}W^{(2m+1)}\rightarrow H^{(2(l+1))}$ \\
\hline 
 & $Z^{(2n)}\rightarrow Z^{(2m)}H^{(2(l+1))}$ &  \\
Associated $H^{(n)}$ & $Z^{(2n)}\rightarrow Z^{(2m+1)}H^{(2l+1)}$ & $Z^{(0)}\rightarrow Z^{(2m)}H^{(2(l+1))}$ \\
Production & $Z^{(2n+1)}\rightarrow Z^{(2m)}H^{(2l+1)}$ & $Z^{(0)}\rightarrow Z^{(2m+1)}H^{(2l+1)}$ \\
 & $Z^{(2n+1)}\rightarrow Z^{(2m+1)}H^{(2(l+1))}$ & \\
\hline
KK parity & $\times$ & $\bigcirc$ \\
\hline
Dark matter & $\times$ & $\bigcirc$ \\
\hline
\hline
 & Our Model  \\
\cline{1-2}
\cline{1-2}
Top Yukawa & \Gc{2}{$10\%$} \\
deviation &\\
\cline{1-2}
$H^{(0)}$ production & \Gc{2}{$81\%$ of SM} \\
by $gg$ fusion &\\
\cline{1-2}
$H^{(0)}$ production & \Gc{2}{$81\%$ of SM} \\
by $WW$ fusion & \\
\cline{1-2}
& \hspace{-0.6cm}$q^{(2n)}\bar{q}^{(2|n-l-1|)}\rightarrow H^{(2(l+1))}$ \\
$H^{(n)}$ production & \hspace{-0.85cm}$q^{(2n)}\bar{q}^{(2|n-l|+1)}\rightarrow H^{(2l+1)}$ \\
by $gg$ fusion & \hspace{-0.85cm}$q^{(2n+1)}\bar{q}^{(2|n-l|)}\rightarrow H^{(2l+1)}$ \\
 & $q^{(2n+1)}\bar{q}^{(|2(n-l)-1|)}\rightarrow H^{(2(l+1))}$ \\
\cline{1-2} 
 & \hspace{-6mm}$W^{(2n)}W^{(2|n-l-1|)}\rightarrow H^{(2(l+1))}$ \\
$H^{(n)}$ production & \hspace{-0.8cm}$W^{(2n)}W^{(2|n-l|+1)}\rightarrow H^{(2l+1)}$ \\
by $WW$ fusion & \hspace{-0.45cm}$W^{(2n+1)}W^{(2|n-l|+1)}\rightarrow H^{(2l+1)}$ \\
 & $W^{(2n+1)}W^{(|2(n-l)-1|)}\rightarrow H^{(2(l+1))}$ \\
\cline{1-2}
Associated $H^{(n)}$ & \Gc{2}{$Z^{(0)}\rightarrow Z^{(l)}H^{(l)}$} \\
Production & \\
\cline{1-2}
KK parity & $\bigcirc$ \\
\cline{1-2}
Dark matter & $\bigcirc$ \\
\cline{1-2}
\end{tabular}
\end{center}
\caption{
Comparisons of phenomenological consequences of our model with BLP models 
presented in \cite{Haba:2009uu}. Our model and BLP(BF) model have a reflection 
symmetry. "81\% of SM" means the ratio of the production cross section. And, 
the $\hat{\lambda}$ is a quartic coupling of Higgs in the brane localized 
potential.
}
\label{tab3}
\end{table}

Finally, we comment on the KK masses including the radiative 
corrections and LKP as DM candidate in the BLP(BF) model. The radiative 
corrections for the neutral KK Higgs~\eqref{KKH} can be estimated as
 \begin{eqnarray}
  \delta m_{H^{(n)}}^2
   \simeq\frac{n^2m_{KK}^2}{16\pi^2}
         \left(3g^2+\frac{3}{2}g'{}^2-\hat{\lambda}\right)\ln
         \frac{\Lambda}{nm_{KK}}+\bar{m}_H^2,
 \end{eqnarray}
where $\hat{\lambda}$ and $\bar{m}_H$ indicate the Higgs quartic coupling and 
boundary mass term. Since the mass of KK photon including the radiative 
correction is the same as one of the UED and the model discussed in this paper,
 the KK photon becomes the LKP as the DM candidate in the BLP(BF) model as long
 as the reliable Higgs quartic coupling and boundary mass term are 
taken.\footnote{See \cite{Haba:2009uu} for detailed descriptions of zero-mode 
Higgs mass and boundary mass term. And see also \cite{Haba:2009wa} for the (KK)
 Higgs masses in a case that the boundary Higgs potentials can be 
perturbatively dealt with.}

\section{Summary}

We have studied the 5D model where the all SM fields exist in the 5D bulk 
compactified on the line segment with the flat metric. The wave-function 
profiles of the Higgs and gauge fields have been presented under the Dirichlet 
BC for the Higgs and the Neumann BC for the gauge fields on the branes. It has 
been shown that a sufficiently flat profile of the longitudinal component of the
 zero-mode gauge boson can be obtained by a superposition of the KK-mode of NG 
bosons.

We have also presented phenomenological discussions on the top Yukawa 
deviation, the production and decay of Higgs, constraints on the model from the
 EW precision measurements, and dark matter candidate. In the model, the top 
Yukawa deviation of 10\% is predicted. The cross section for the Higgs 
production, such as the gluon and $WW$ fusions, in this model are over all 
decreased to 81\% of the SM expectations. Therefore, the branching ratios are 
not changed. For the Higgs decay, the width of the process $H\rightarrow WW$, 
which is the dominant one, is as large as the Higgs mass. The evaluation of the 
$S$ and $T$ parameters suggests that the KK scale of $430\mbox{ GeV}\lesssim 
m_H\lesssim 500\mbox{ GeV}$ is favored in this scenario.

\subsection*{Acknowledgments}

We are grateful to T.~Yamashita for very helpful discussions and 
thank S. Matsumoto for useful comments. This work is partially 
supported by Scientific Grant by Ministry of Education and Science, 
Nos.\ 20540272, 20039006, 20025004, 20244028, and 19740171. The work 
of RT is supported by the DFG-SFB TR 27.

\appendix

\section*{Appendix}
\section*{Gauge Fixing}
We give our notation with 5D Higgs kinetic Lagrangian and gauge 
fixing in this Appendix. In our notation, the VEV and quantum 
fluctuation are given by
 \begin{eqnarray}
  \Phi^c(x,z) &=&
   \left(
    \begin{array}{c}
     0 \\
     v
    \end{array}
   \right), \nonumber \\
  \Phi^q(x,z) &=&
   \left(
    \begin{array}{c} 
     \varphi^+(x,z) \\
     \frac{1}{\sqrt{2}}[H(x,z)+i\chi(x,z)]
    \end{array}
   \right)=
   \left(
    \begin{array}{c} 
     \displaystyle\sum_{n=0}^\infty f_n^{\varphi^+}(z)\varphi_n^+(x) 
     \\
     \frac{1}{\sqrt{2}}\displaystyle\sum_{n=0}^\infty
     [f_n^H(z)H_n(x)+if_n^\chi(z)\chi_n(x)]
    \end{array}
   \right). \nonumber 
 \end{eqnarray}
The covariant derivative on the Higgs field is written down as 
 \begin{eqnarray}
  D_M\Phi &=& \partial_M\Phi
              +\frac{ig_5}{\sqrt{2}}
                \left(
                 \begin{array}{cc}
                  0     & W_M^+ \\
                  W_M^- & 0
                 \end{array}
                \right)\Phi
              +ie\left(
                  \begin{array}{cc}
                   \frac{1}{\tan2\theta_W}Z_M+A_M & 0 \\
                   0                              & 
               -\frac{1}{\sin2\theta_W}Z_M
                  \end{array}
                 \right)\Phi \nonumber \\
          &=& \left(
               \begin{array}{c}
                \partial_M\varphi^+                          \\
                \frac{\partial_MH+i\partial_M\chi}{\sqrt{2}}
               \end{array} 
              \right)
              +\frac{ig_5}{\sqrt{2}}
               \left(
                \begin{array}{c}
                 W_M^+\left(v+\frac{H+i\chi}{\sqrt{2}}\right) \\
                 W_M^-\varphi^+
                \end{array}
               \right)
           +ie_5\left(
                  \begin{array}{c}
                   \left(\frac{1}{\tan2\theta_W}Z_M+A_M\right)
                   \varphi^+ \\
                   -\frac{1}{\sin2\theta_W}Z_M
                    \left(v+\frac{H+i\chi}{\sqrt{2}}\right)
                  \end{array}
                 \right), \nonumber 
 \end{eqnarray}
where the gauge fields are defined in Eq.\eqref{gauge}, and 
$	e_5	\equiv	g_5g_5'/\sqrt{g_5^2+{g_5'}^2}$. Note that 
mass dimensions are $[g_5]=[g_5']=[e_5]=-1/2$ and 
$[v]=[W^\pm_M]=[Z_M]=[A_M]=3/2$. The Higgs kinetic Lagrangian is 
given by
 \begin{align}
	\mathcal{L}_H
		& =	-\left|D_M\Phi\right|^2 \nonumber \\
		&=	-\left|
		 		\partial_M\varphi^+
				+im_WW_M^+
				+{ig_5\over2}W_M^+\left(H+i\chi
                                                  \right)
				+ie_5\left({1\over\tan2\theta_W}Z_M
                                +A_M\right)\varphi^+
				\right|^2\nonumber\\
		&\quad
			-{1\over2}\left|
                         \partial_MH+i\partial_M\chi
				+ig_5W_M^-\varphi^+
				-im_ZZ_M
				-{ie_5\over\sin2\theta_W}Z_M
                                 \left(H+i\chi\right)
			\right|^2. \nonumber 
  \end{align}
The quadratic terms are given by 
\begin{align}
	\mathcal{L}_H^{\text{quad}}
		&=	-\left|\partial_M\varphi^+\right|^2
			-{\left(\partial_MH\right)^2
                        +\left(\partial_M\chi\right)^2\over2}
			-m_W^2\left|W_M^+\right|^2
			-{m_Z^2\over2}(Z_M)^2\nonumber\\
		&\quad
			+im_W\left(
				W^{-M}\partial_M\varphi^+
				-W^{+M}\partial_M\varphi^-
				\right)
			+m_ZZ^M\partial_M\chi. \nonumber
  \end{align} 

We employ the following $R_\xi$-$like$ gauge fixing
 \begin{align}
	\mathcal{L}_{\text{GF}}=
   -f^+f^--\frac{1}{2}(f^Zf^Z+f^Af^A), \nonumber 
 \end{align}
where
 \begin{align}
	f^\pm	&\equiv	\frac{1}{\sqrt{\xi_W}}
                        \left(\partial_M W^{\pm M}
			 	\mp i\xi_W m_W\varphi^\pm\right), 
                        \nonumber \\
	f^Z	&\equiv \frac{1}{\sqrt{\xi_Z}}\left(\partial_M Z^M
				-\xi_Z m_Z\chi\right), \nonumber \\
	f^A	&\equiv	
                 \frac{1}{\sqrt{\xi_A}}\partial_M A^M.\nonumber 
 \end{align}
Now, we can write down as
\begin{align}
	\mathcal{L}_{\text{GF}}
		&=	-{1\over\xi_W}\left|\partial_MW^{+M}\right|^2
			-{1\over2\xi_Z}
				\left(\partial_MZ^M\right)^2
			-{1\over2\xi_A}\left(\partial_MA^M\right)^2
				\nonumber\\
		&\quad
			+im_W\left(\varphi^+\partial_MW^{-M}
                        -\varphi^-\partial_MW^{+M}\right)
			+m_Z\chi\partial_MZ^M
				-\xi_W m_W^2\left|\varphi^+\right|^2
				-{\xi_Z m_Z^2\over 2}\chi^2.
   \nonumber 
  \end{align}
The sum of quadratic terms from Higgs kinetic Lagrangian and gauge 
fixing terms are given by
\begin{align}
	\mathcal{L}_{H+\text{GF}}^{\text{quad}}
		&=	-\left|\partial_M\varphi^+\right|^2
			-\xi_Wm_W^2\left|\varphi^+\right|^2
			-{1\over2}\left(\partial_M\chi\right)^2
			-{\xi_Zm_Z^2\over2}\chi^2
			-{1\over2}\left(\partial_Mh\right)^2
                  \nonumber\\
		&\quad
			-\frac{1}{\xi_W}\left|\partial_NW^{+N}
                         \right|^2
			-m_W^2\left|W_M^+\right|^2
			-{1\over2\xi_Z}\left(\partial_NZ^N\right)^2
			-{m_Z^2\over2}(Z_M)^2
			-{1\over2\xi_A}\left(\partial_MA^M\right)^2
                 \nonumber\\
		&\quad
			+\partial_5\left[im_W\left(W^-_5\varphi^+
                        -W^+_5\varphi^-\right)+m_ZZ_5\chi\right].
                 \nonumber			
				\label{quad_terms}
 \end{align}
One can consider some specific gauge choices. Here, we comment on the
 't Hooft-Feynmann gauge ($\xi_V=1$, $V=W,Z,A$) and unitary gauge 
($\xi_V\rightarrow\infty$). The above sum of quadratic terms can be 
simplified to be 
 \begin{align}
	\mathcal{L}_{H+\text{GF}_{\xi_V=1}}^{\text{quad}}
		&=	-\left|\partial_M\varphi^+\right|^2
			-m_W^2\left|\varphi^+\right|^2
			-{1\over2}\left(\partial_M\chi\right)^2
			-{m_Z^2\over2}\chi^2
			-{1\over2}\left(\partial_Mh\right)^2
                  \nonumber\\
		&\quad
			-\left|\partial_NW^{+N}\right|^2
			-m_W^2\left|W_M^+\right|^2
			-{1\over2}\left(\partial_NZ^N\right)^2
			-{m_Z^2\over2}(Z_M)^2
			-{1\over2}\left(\partial_MA^M\right)^2
                  \nonumber\\
		&\quad
	         +\partial_5\left[im_W\left(W^-_5\varphi^+
                 -W^+_5\varphi^-\right)+m_ZZ_5\chi\right]. \nonumber 
 \end{align} 
in the 't Hooft-Feynmann gauge, and 
 \begin{align}
  \mathcal{L}_{H+\text{GF}_{\xi_V\rightarrow\infty}}^{\text{quad}}
		\rightarrow	
			-{1\over2}\left(\partial_Mh\right)^2
			-m_W^2\left|W_M^+\right|^2
			-{m_Z^2\over2}(Z_M)^2, \nonumber 
 \end{align}
in the unitary gauge. Notice that unphysical degrees of freedom, that
 is, the would-be NG bosons $\varphi^\pm$ and $\chi$, become 
infinitely heavy and decouple.

Finally, let us give the gauge kinetic Lagrangian in the 
't Hooft-Feynmann gauge. The gauge kinetic Lagrangian is written by
 \begin{align}
	\mathcal{L}_{\text{YM}}
		&=
  -{1\over4}\left(\sum_{a=1}^3F^a_{MN}F^{aMN}
  +F^B_{MN}F^{B\,MN}\right). \nonumber 
 \end{align}
By utilizing the following redefinition,
 \begin{align}
	F^\pm_{MN}
		&=	\partial_MW^\pm_N-\partial_NW^\pm_M
			\pm 2ig\left(W_M^3W^\pm_N-W_N^3W_M^\pm
                               \right),\nonumber \\
	F^3_{MN}
		&=	\partial_MW^3_N-\partial_NW^3_M
			+2g\left(W^+_MW^-_N-W^+_NW^-_M\right),
                 \nonumber 
 \end{align}
with $W^3_M=cZ_M+sA_M$, we can rewrite
 \begin{align}
	\mathcal{L}_{\text{YM}}
		&=	-{1\over2}F^+_{MN}F^{-MN}
			-{1\over4}\left[
				F^3_{MN}F^{3\,MN}+F^B_{MN}F^{B\,MN}
                 \right].\nonumber 
  \end{align}
Therefore, quadratic terms are\footnote{We do not consider 
Wilson-line phases and put all the VEVs of gauge field zero.}
\begin{align}
	\mathcal{L}_{\text{YM}}^{\text{quad}}
		&=	-{1\over2}\sum_{a=1}^3\left(
				-W^a_\mu\Box W^{a\mu}
				-\left(\partial_\mu W^{a\mu}\right)^2
				+\left(\partial_5W^a_\mu\right)
                                 \left(\partial_5W^{a\mu}
					\right)
				-W_5^a\Box W_5^a\right. \nonumber\\
		&\left.\hspace{9.9cm}		
                 +2W_5^a\partial_5\left(\partial_\mu W^{a\mu}\right)
				\right)\nonumber\\
		&\quad
			-{1\over2}\left(
				-B_\mu\Box B^{\mu}
				-\left(\partial_\mu B^{\mu}\right)^2
				+\left(\partial_5B_\mu\right)
                                 \left(\partial_5B^{\mu}\right)
				-B_5\Box B_5
				+2B_5\partial_5
                                 \left(\partial_\mu B^{\mu}\right)
				\right)\nonumber\\
		&=	-\left[
				-W^+_\mu\Box W^{-\mu}
				-\left|\partial_\mu W^{+\mu}\right|^2
				+\left(\partial_5 W^+_\mu\right)
                                 \left(\partial_5 W^{-\mu}\right)
				\right] \nonumber\\
		&\quad
			-\left[
				-W_5^+\Box W_5^-
				+W_5^+\partial_5
                                 \left(\partial_\mu W^{-\mu}\right)
				+W_5^-\partial_5
                                 \left(\partial_\mu W^{+\mu}\right)
				\right]\nonumber\\
		&\quad
			-{1\over2}\left(
				-Z_\mu\Box Z^{\mu}
				-\left(\partial_\mu Z^{\mu}\right)^2
				+\left(\partial_5Z_\mu\right)
                                 \left(\partial_5Z^{\mu}\right)
				-Z_5\Box Z_5
				+2Z_5\partial_5
                                 \left(\partial_\mu Z^{\mu}\right)
				\right)\nonumber\\
		&\quad
			-{1\over2}\left(
				-A_\mu\Box A^{\mu}
				-\left(\partial_\mu A^{\mu}\right)^2
				+\left(\partial_5A_\mu\right)
                                 \left(\partial_5A^{\mu}\right)
				-A_5\Box A_5
				+2A_5\partial_5
                                 \left(\partial_\mu A^{\mu}\right)
				\right). \nonumber 
  \end{align}

\end{document}